\documentstyle{ichep}
\include{mytex}
\begin{document}
\newcommand{\dis}{\displaystyle}
\title{{\LARGE{\bf Schemes for Neutrino Mass and Mixing}}}
\author{ {\large {\bf Jos\'e W. F. Valle $^{\dag}$ } }}
\address{\dag\
Departament de F\'{\i}sica Te\'{o}rica, Universitat de Val\`{e}ncia
and \\ Instituto de F\'{\i}sica Corpuscular - C.S.I.C.,
E-46100 Burjassot, Val\`encia, SPAIN         }
\abstract
{I briefly review various schemes of \neu mass
generation which are motivated by present experimental
hints from solar and atmospheric neutrinos as well as
cosmological data on the amplitude of primordial
density fluctuations. }

\twocolumn[\maketitle]

\section{Preliminaries}
\hspace{0.5cm}

Neutrinos are the only apparently massless electrically neutral
fermions in the \sm and the only ones without \rh partners.
It is rather mysterious that they seem to be so special
when compared with the other fundamental fermions.
Indeed, having no electric charge, a majorana mass
term for \neus may arise even in the absence of \rh
components. On the other hand, many unified extensions
of the \sm, such as SO(10), do require the presence of
\rh neutrinos in order to realize the extra symmetry.
Either way one expects \neus to be massive. Moreover,
there is, in these theories, a natural mechanism,
called seesaw, to understand the relative smallness
of \neu masses \cite{GRS,fae}. In general the seesaw
mechanism provides just a general scheme, rather than
detailed predictions. These will depend, among other
factors, upon the structure not only of the Dirac type
entries, but also on the possible texture of the large
Majorana mass term \cite{Smirnov}.

Although attractive, the seesaw mechanism is by no means
the only way to generate \neu masses. There are many other
attractive possibilities, some of which do not require any
new large mass scale. The extra particles required to
generate the \neu masses have masses at scales accessible
to present experiments \cite{zee.Babu88}.

It is also quite plausible that B-L or lepton number,
instead of being part of the \gau symmetry \cite{LR}
may be a spontaneously broken global symmetry. The scale
at which such a symmetry gets broken does not need to high,
as in the original proposal \cite{CMP}, but can be rather
low, close to the weak scale \cite{JoshipuraValle92}.
Such a low scale for lepton number breaking could have
important implications not only in astrophysics and
cosmology but also in particle physics.

Unfortunately, present theory is not capable of
predicting the scale of \neu masses any better
than it can fix the masses of the other fermions,
say the muon. One should at this point turn
to experiment.

There are several limits on \neu masses that follow
from observation. The laboratory bounds may be
summarized as \cite{PDG94}
\beq
\label{1}
m_{\nu_e} 	\lsim 5 \: \rm{eV}, \:\:\:\:\:
m_{\nu_\mu}	\lsim 250 \: \rm{keV}, \:\:\:\:\:
m_{\nu_\tau}	\lsim 31  \: \rm{MeV}
\eeq
and follow purely from kinematics. These are the
most model-independent of the \neu mass limits.
The improved limit on the \ne mass was given by
Lobashev at this conference \cite{Erice}, while
that on the \nt mass may be substantially
improved at a future tau factory \cite{jj}.

In addition, there are limits on neutrino masses
that follow from the nonobservation of neutrino
oscillations \cite{granadaosc}, which involve
\neu mass differences versus mixing, and disappear
in the limit of unmixed neutrinos.

Another important limit arises from the
non-observation of ${\beta \beta}_{0\nu}$ decay, i.e.
the process by which nucleus $(A,Z-2)$ decays to
$(A,Z) + 2 \ e^-$. This lepton number violating
process would arise from majorana \neu exchange.
In fact, as shown in ref. \cite{BOX}, a nonvanishing
${\beta \beta}_{0\nu}$ decay rate requires \neus to be
majorana particles, irrespective of which mechanism
induces it. This establishes a very deep connection
which, in some special models, may be translated
into a lower limit on the \neu masses.
The negative searches for ${\beta \beta}_{0\nu}$
in $^{76} \rm{Ge}$ and other nuclei leads to a limit
of about one or two eV
on a weighted average \neu mass parameter
characterizing this process. Better sensitivity is expected
from the upcoming enriched germanium experiments.
Although rather stringent, this limit may allow relatively
large \neu masses, as there
may be strong cancellations between the contributions of
different neutrino types. This happens automatically
in the case of a Dirac \neu due to the lepton number
symmetry \cite{QDN}.


In addition to laboratory limits, there is a cosmological
bound that follows from avoiding the overabundance of
relic neutrinos \cite{KT}
\beq
\sum_i m_{\nu_i} \lsim 50 \: \rm{eV}
\label{rho1}
\eeq
This limit only holds if \neus are stable on cosmological
time scales. There are many models where neutrinos decay
into a lighter \neu plus a majoron \cite{fae},
\beq
\nu_\tau \ra \nu_\mu + J \:\: .
\label{NUJ}
\eeq
Lifetime estimates in various majoron models have
been discussed in ref. \cite{V}. These decays can
be fast enough to obey the cosmological limits coming
from the critical density requirement, as well as those
that come from primordial big-bang nucleosynthesis
\cite{BBNUTAU}. Note also that, since these decays
are $invisible$, they are consistent with all
astrophysical observations.
In view of the above it is worthwhile to continue
in the efforts to improve present laboratory \neu mass
limits, including searches for distortions in the energy
distribution of the electrons and muons coming from weak
decays \sa $\pi, K \ra e \nu$, $\pi, K \ra \mu \nu$, as
well as kinks in nuclear $\beta$ decays \cite{Deutsch}.

\section{Hints for Neutrino Mass}
\hspace{0.5cm}

In addition to the above limits there are
some positive {\sl hints} for neutrino masses
that follow from the following cosmological,
astrophysical and laboratory observations.

\subsection{Dark Matter}
\hspace{0.5cm}

Recent observations of cosmic background temperature
anisotropies on large scales by the COBE  satellite
\cite{cobe}, when combined with cluster-cluster correlation
data e.g. from IRAS \cite{iras}, indicate the need for the existence
of a hot {\sl dark matter} component, contributing
about 30\% to the total mass density \cite{cobe2}.
A good fit is provided by a massive neutrino, \sa as a
\nt of a few eV mass. This suggests the possibility
of having observable \ne to \nt or \nm to \nt
oscillations that may be accessible to the CHORUS
and NOMAD experiments at CERN, as well as at the
proposed P803 experiment at Fermilab \cite{chorus}.
This mass scale is also consistent with the recent
hints reported here by Caldwell \cite{Caldwell}.

\subsection{Solar Neutrinos}
\hspace{0.5cm}

The data collected up to now by
the two high-energy experiments Homestake and Kamiokande,
as well as by the low-energy data on pp neutrinos from
the GALLEX and SAGE experiments still pose a persisting
puzzle \cite{Davis,granadasol}.

Comparing the data of GALLEX with the Kamiokande data
indicates the need for a reduction of the $^7 $ Be flux
relative to the standard solar model expectation. Inclusion
of the Homestake data only aggravates the discrepancy,
suggesting that the solar \neu problem is indeed a real problem.

The simplest astrophysical solutions to the solar \neu data
are highly disfavored \cite{NEEDNEWPHYSICS}.
The most attractive way to account for the data
is to assume the existence of \neu conversions
involving very small \neu masses $\sim 10^{-3}$ eV
\cite{MSW}. The region of parameters allowed by present
experiments is given in ref. \cite{Hata.MSWPLOT}.
Note that the fits favour the non-adiabatic over the
large mixing solution, due mostly to the larger reduction
of the $^7 $ Be flux found in the former.

\subsection{Atmospheric Neutrinos}
\hspace{0.5cm}

An apparent decrease in the expected flux of atmospheric
$\nu_\mu$'s relative to $\nu_e$'s arising from the decays
of $\pi$'s, $K$'s and secondary muon decays produced in
the atmosphere, has been observed in two underground
experiments, Kamiokande and IMB, and possibly also at
Soudan2 \cite{atm}. This atmospheric neutrino deficit
can be ascribed to \neu oscillations.
Although the predicted absolute
fluxes of \neus produced by cosmic-ray interactions in the
atmosphere are uncertain at the 20 \% level, their
ratios are expected to be accurate to within
5 \%.

Combining these experimental results with observations
of upward going muons made by Kamiokande, IMB and Baksan,
and with the negative Frejus and NUSEX results \cite{up}
leads to the following range of neutrino oscillation
parameters \cite{atmsasso}
\beq
\label{atm0}
\Delta m^2_{\mu \tau} \approx 0.005 \: - \: 0.5\ \rm{eV}^2,\
\sin^22\theta_{\mu \tau} \approx 0.5
\eeq
Recent results from Kamiokande on higher energy \neus
strengthen the case for an atmospheric \neu problem
\cite{atm1}.

\section{Models Reconciling Present Hints.}
\hspace{0.5cm}

Can we reconcile the present hints from astrophysics and
cosmology in the framework of a consistent elementary
particle physics theory? The above observations suggest
an interesting theoretical puzzle whose possible
resolutions I now discuss.

\subsection{Three Almost Degenerate Neutrinos}
\hspace{0.5cm}

It is difficult to reconcile these three observations
simultaneously in the framework of the simplest seesaw model
with just the three known \neus. The only possibility is if
all three \neus are closely degenerate \cite{caldwell}.

It is known that the general seesaw models have
two independent terms giving rise to the light \neu masses.
The first is an effective triplet vacuum expectation value
\cite{2227} which is expected to be small in left-right
symmetric models \cite{LR}. Based on this fact one can
in fact construct extended seesaw models where the main
contribution to the light \neu masses ($\sim$ 2 eV) is universal,
due to a suitable horizontal symmetry, while the splittings
between \ne and \nm explain the solar \neu deficit and that
between \nm and \nt explain the atmospheric \neu anomaly \cite{DEG}.

\subsection{Three Active plus One Sterile Neutrino}
\hspace{0.5cm}

The alternative way to fit all the data is to add a
fourth \neu species which, from the LEP data on the
invisible Z width, we know must be of the sterile type,
call it \ns. The first scheme of this type gives mass
to only one of the three neutrinos at the tree level,
keeping the other two massless \cite{OLD}.
In a seesaw scheme with broken lepton number, radiative
corrections involving gauge boson exchanges will give
small masses to the other two neutrinos \ne and \nm
\cite{Choudhury}. However, since the singlet \neu is
superheavy in this case, there is no room to account
for the three hints discussed above.

Two basic schemes have been suggested to reconcile all
three hints. In addition to a light sterile \neu \ns, they
invoke additional Higgs bosons beyond that of the standard
model. In these models the \ns either lies at the dark matter
scale \cite{DARK92} or, alternatively, at the solar \neu scale
\cite{DARK92B}.
In the first case the atmospheric
\neu puzzle is explained by \nm to \ns oscillations,
while in the second it is explained by \nm to \nt
oscillations. Correspondingly, the deficit of
solar \neus is explained in the first case
by \ne to \nt oscillations, while in the second
it is explained by \ne to \ns oscillations. In both
cases it is possible to fit all observations together.
However, in the first case there is a clash with the
bounds from big-bang nucleosynthesis. In the latter
case the \ns is at the MSW scale so that nucleosynthesis
limits are satisfied. They single out the nonadiabatic
solution uniquely. Note however that, since the
mixing angle characterizing the \nm to \nt
oscillations is nearly maximal, the second
solution is in apparent conflict with \eq{atm0}
but agrees with Fig. 5 of ref. \cite{atm1}.
Another theoretical possibility is that all active
\neus are very light, while the sterile \neu \ns is
the single \neu responsible for the dark matter
\cite{DARK92D}.

\section{Outlook}
\hspace{0.5cm}

Besides being suggested by theory, neutrino masses
seem to be required to fit present astrophysical and
cosmological observations, in addition to the recent
LSND hints discussed here \cite{Caldwell}.

Neutrinos could be responsible for a wide variety of
measurable implications at the laboratory. These new
phenomena would cover an impressive range of energies,
starting with $\beta$ and nuclear $\beta \beta_{0\nu}$
decays. Searches for the latter with enriched germanium
could test the quasidegenerate neutrino scenario for
the joint explanation of hot dark matter and
solar and atmospheric \neu anomalies.  Moving to
neutrino oscillations, here one expects much larger
regions of oscillation parameters in the \ne to \nt
and \nm to \nt channels will be be probed by the
accelerator experiments at CERN than now possible
with present accelerators and reactors.
On the other hand more data from low energy pp
neutrinos as well as from Superkamiokande, Borexino,
and Sudbury will shed light on the solar neutrino issue.

For the far future we look forward to the possibility
of probing those regions of \nm to \ne or \ns oscillation
parameters suggested by present atmospheric \neu data.
This will be possible at the next generation of long
baseline experiments.
Similarly, a new generation of experiments capable
of more accurately measuring the cosmological
temperature anisotropies at smaller angular scales than
COBE, would test different models of structure formation,
and presumably shed further light on the need for hot
\neu dark matter.

Neutrinos may also imply rare processes with lepton
flavour violation, as well as new signatures at LEP
energies and even higher. Such experiments may be
complementary to those at low energies and can
also indirectly test \neu properties in an
important way.

\section*{Acknowledgements}
\hspace{0.5cm} This paper has been supported by DGICYT under
Grant number PB92-0084.


\Bibliography{15}

\bibitem{GRS}
 M Gell-Mann, P Ramond, R. Slansky,
in {\sl Supergravity},  ed. D. Freedman et al. (1979);
 T. Yanagida,
 in {\sl KEK lectures},  ed.  O. Sawada et al. (1979)

\bibitem{fae}
For a recent review see
J. W. F. Valle, {\sl Gauge Theories and the Physics of
Neutrino Mass}, \ppnp{26}{91}{91-171} and references therein.

\bibitem{Smirnov}
A. Yu. Smirnov, \pr{D48}{94}{3264};
E. Papageorgiou, paper 807

\bibitem{zee.Babu88}
 A. Zee, \pl{B93}{80}{389};
 K.~S. Babu, \pl{B203}{88}{132}

\bibitem{LR}
R.N.~Mohapatra and G.~Senjanovic, \pr{D23}{81}{165}
and references therein.

\bibitem{CMP}
Y. Chikashige, R. Mohapatra, R. Peccei, \prl{45}{80}{1926}

\bibitem{JoshipuraValle92}
 J. W.~F. Valle, \tp

\bibitem{PDG94}
 Particle Data Group, \pr{D50}{94}{1173}

\bibitem{Erice}
V. Lobashev, \tp

\bibitem{jj}
J. Gomez-Cadenas, M. C. Gonzalez-Garcia, \pr {D39} {89} {1370};
J. Gomez-Cadenas \etal \pr{D41} {90} {2179}

\bibitem{granadaosc}
J Schneps,
\nps{31}{93}{307}.

\bibitem{BOX}
 J. Schechter and  J. W.~F. Valle, \pr{D25}{82}{2951}

\bibitem{QDN}
 J. W.~F. Valle, \pr{D27}{83}{1672} and references therein;
 L. Wolfenstein, \np{B186}{81}{147}

\bibitem{KT}
E. Kolb, M. Turner, {\sl The Early Universe},
Addison-Wesley, 1990.

\bibitem{V}
J. W. F. Valle, \pl {B131} {83}{87};
G. Gelmini, J. W. F. Valle, \pl {B142} {84}{181};
M. C. Gonzalez-Garcia, J. W. F. Valle, \pl {B216} {89} {360}.
 A. Joshipura, S. Rindani, PRL-TH/92-10; for an early discussion
 see  J. Schechter and  J. W. F. Valle,  \pr{D25}{82}{774}

\bibitem{BBNUTAU}
G. Steigman; proceedings of the
{\sl Int. School on Cosmological Dark Matter},
(World Scientific, 1994), ed. J. W. F. Valle and A. Perez, p. 55

\bibitem{Deutsch}
See, e.g. J Deutsch etal \np{A518}{90}{149}; \pw{2}{91}{81};
A. Hime,  \nps{31}{93}{50}.

\bibitem{cobe}
G.~F. Smoot et~al., \apj{396}{92}{L1-L5};
E.L.~Wright et al., \apj{396}{92}{L13}

\bibitem{iras}
R. Rowan-Robinson, proceedings of the {\sl International
School on Cosmological Dark Matter}, \opc p. 7-18

\bibitem{cobe2}
E.L.~Wright et al., \apj{396}{92}{L13};
M.~Davis, F.J.~Summers, and D.~Schagel, \nat{359}{92}{393};
A.N.~Taylor and M.~Rowan-Robinson, \ib{359}{92}{396};
R.K.~Schaefer and Q.~Shafi, \nat{359}{92}{199};
J.A.~Holtzman and J.R.~Primack, \apj{405}{93}{428};
A.~Klypin et al., \apj{416}{93}{1}

\bibitem{chorus}
CHORUS and NOMAD proposals CERN-SPSLC/91-21 (1992) and
CERN-SPSC/90-42 (1992); K. Kodama et~al., FNAL proposal P803 (1991).

\bibitem{Caldwell}
D. Caldwell, \tp

\bibitem{Davis}
 J. R.~Davis
 in {\em Proceedings of the 21th International Cosmic Ray Conference,
  Vol. 12},  ed.\  R.~J. Protheroe (University of Adelaide Press, 1990)
  p. 143.

\bibitem{granadasol}
GALLEX collaboration,
\pl{B285}{92}{376}, \ib{B285}{92}{390};
\pl{B314}{93}{445}; \pl{B327}{94}{377}  

\bibitem{NEEDNEWPHYSICS}
J. Bahcall, H. Bethe, \pr{D47}{93}{1298}, \prl{65}{90}{2233};
V. Berezinsky, LNGS-93/86;
S. Bludman, N. Hata, P. Langacker,
\pr{D45}{92}{1820}; X. Shi, D. Schramm, FERMILAB-PUB-92-322-A

\bibitem{MSW}
M. Mikheyev, A. Smirnov, \sjnp{42}{86}{913};
L. Wolfenstein, \pr {D17}{78}{2369}; \ib{D20}{79}{2634}.

\bibitem{Hata.MSWPLOT}
N. Hata, P. Langacker, \pr{D50}{94}{632} and Pennsylvania
preprints UPR-0592-T and UPR-0625-T and references therein

\bibitem{atm}
Kamiokande collaboration, \pl{B205}{88}{416};
\pl{B280}{92}{146} and \pl{B283}{92}{446} ;
IMB collaboration, \pr{D46}{92}{3720}

\bibitem{atmsasso}
Proceedings of {\sl Int. Workshop on \nm/\ne problem
in atmospheric \neus} ed. V. Berezinsky and G Fiorentini,
Gran Sasso, 1993.

\bibitem{up}
M.M.~Boliev et al.\ in Proceedings of the 3rd International
Workshop on Neutrino Telescopes, p.~235;
%
Ch.~Berger et al., \pl{B245}{90}{305}; \ib{227}{89}{489}; 
M.~Aglietta et al., \jel{15}{91}{559}. 

\bibitem{atm1}
Kamiokande collaboration, preprint ISSN 1340-3745

\bibitem{caldwell}
D.O.~Caldwell and R.N.~Mohapatra, \pr{D48}{93}{3259};
A. S. Joshipura, preprint  PRL-TH/93/20, December 1993,
to appear in \zp{}{94}{};
S. T. Petcov, A. Smirnov, \pl{B322}{94}{109}

\bibitem{2227}
 J. Schechter and  J. W. F. Valle, \pr{D22}{80}{2227}

\bibitem{DEG}
A. Ioannissyan, J.W.F. Valle, \pl{B332}{94}{93-99};
D.O.~Caldwell and R.N.~Mohapatra, preprint UCSB-HEP-94-03;
B. Bamert, C.P. Burgess, preprint McGill-94/07;
D. Caldwell and R. N. Mohapatra, Maryland report, UMD-PP-94-90 (1994);
D. G. Lee and R. N. Mohapatra, Maryland Report, UMD-PP-94-95 (1994);
A. S. Joshipura, preprint  PRL-TH/94/08

\bibitem{OLD}
 J. Schechter and  J. W. F. Valle, \pr{D21}{80}{309}

\bibitem{Choudhury}
D. Choudhury et al \pr{D50}{94}{3486}; similar
model was considered by G. Hou, Wong, paper 703

\bibitem{DARK92}
J.~T. Peltoniemi, D.~Tommasini, and J~W~F Valle,
\pl {B298}{93}{383}

\bibitem{DARK92B}
J.~T. Peltoniemi, and J~W~F Valle, \np{B406}{93}{409};
E. Akhmedov, Z. Berezhiani, G. Senjanovic and Z. Tao,
\pr{D47}{93}{3245}.

\bibitem{DARK92D}
J.~T. Peltoniemi, \mpl{A38}{93}{3593}

\end{thebibliography}
\end{document}